\documentclass{ws-ijmpcs}

\begin{document}

\markboth{Felix Wick}
{Hidden Charm Spectroscopy from Tevatron}

%
\catchline{}{}{}{}{}
%

\title{HIDDEN CHARM SPECTROSCOPY FROM TEVATRON}

\author{FELIX WICK\footnote{On behalf of the CDF Collaboration.}}

\address{Institut fuer Experimentelle Kernphysik, Karlsruhe Institute of Technology\\
Wolfgang-Gaede-Str. 1, Karlsruhe, 76131, Germany\\
wick@fnal.gov}

\maketitle


\begin{abstract}
  The observation of a narrow structure near the $J/\psi\,\phi$
  threshold in exclusive $B^+ \to J/\psi\,\phi\,K^+$ decays produced
  in $p\bar{p}$ collisions at $\sqrt{s} = 1.96\,\mathrm{TeV}$ is
  reported. A signal of $19 \pm 6(\mathrm{stat}) \pm 3(\mathrm{syst})$
  events, with statistical significance of $5.0$ standard deviations,
  is seen in a data sample corresponding to an integrated luminosity
  of $6.0\,\mathrm{fb}^{-1}$, collected by the CDF II detector. The
  mass and natural width of the structure are determined to be $m =
  4143.4^{+2.9}_{-3.0}(\mathrm{stat}) \pm
  0.6(\mathrm{syst})\,\mathrm{MeV}/c^2$ and $\Gamma =
  15.3^{+10.4}_{-6.1}(\mathrm{stat}) \pm
  2.5(\mathrm{syst})\,\mathrm{MeV}/c^2$, consistent with the previous
  measurements reported as evidence of the $Y(4140)$.
\end{abstract}

\ccode{PACS numbers: 14.40.Rt, 13.25.Gv}

\section{Introduction}

In recent years, several states with charmonium-like decays were
discovered, which do not fit properly into the established charmonium
picture. These states, called $X,Y,Z$, are candidates for exotic
mesons beyond the conventional quark-antiquark model ($q\bar{q}$).
Possible interpretations are quark-gluon hybrids ($q\bar{q}g$),
four-quark states ($q\bar{q}q\bar{q}$), molecular states composed of
two usual mesons, glueballs, et cetera.

CDF reported an evidence for a narrow near-threshold structure in the
$J/\psi \phi$ mass spectrum, called $Y(4140)$, using exclusive $B^+
\to J/\psi \phi K^+$ decays.\cite{Evidence} As there was no signal
seen by Belle in a subsequent search,\cite{Belle} it is important to
investigate with a larger CDF data sample. Consisting of two vector
mesons (positive C-parity), the final state $J/\psi \phi$ is a good
channel to search for an exotic meson. The observed excess would be
the first charmonium-like structure decaying into two heavy quarkonium
states ($c\bar{c}$ and $s\bar{s}$). Because the mass in this channel
is high enough for open charm decays, an explanation as charmonium
state is very unlikely for a narrow structure. A search near the
$J/\psi\,\phi$ threshold is also motivated by the closeness of the
$Y(3930)$ to the $J/\psi\,\omega$ threshold. Some of the possible
exotic explanations for the $Y(4140)$ are discussed in
Ref.~\refcite{Theo}.

\section{Candidate Selection}

We report on an update of the search for structures in the $J/\psi
\phi$ system produced in exclusive $B^+ \to J/\psi \phi K^+$
decays.\cite{PublicNote} The employed dataset was collected by the CDF
II detector at the Tevatron and corresponds to an integrated
luminosity of $6.0\,\mathrm{fb^{-1}}$. It was accumulated using a
dedicated dimuon trigger which requires a $\mu^+ \mu^-$ pair with a
mass of $2.7 < m(\mu^+ \mu^-) < 4.0\,\mathrm{GeV}/c^2$. Due to trigger
prescales at increasing instantaneous luminosities one cannot expect a
linear increase of the sample size compared to the previous
analysis\cite{Evidence} with $2.7\,\mathrm{fb^{-1}}$.

In order to build $B^+ \to J/\psi \phi K^+$ candidates, first $J/\psi
\to \mu^+ \mu^-$ and $\phi \to K^+ K^-$ candidates are reconstructed
which are then combined with an additional charged track with kaon
mass hypothesis. Thereby, the reconstructed $J/\psi$ and $\phi$ masses
are required to lie within $50\,\mathrm{MeV/c^2}$ ($J/\psi$)
respective $7\,\mathrm{MeV/c^2}$ ($\phi$) of the corresponding world
average values. The combinatorial background can be reduced
significantly with a higher threshold of the $B^+$ decay length in the
transverse plane ($L_{xy}(B^+)$) to exploit the long $B$ meson
lifetime. In addition, a kaon identification quantity can be used to
obtain a further background reduction by separating the final state
kaons from the dominant pion background. For that purpose, the
information about the ionization energy loss $dE$/$dx$ in the drift
chamber and the information from the Time-of-Flight detector are
combined in a log-likelihood ratio $LLR_\mathrm{Kaon}$. The cuts
$L_{xy}(B^+)>500\,\mu\mathrm{m}$ and $LLR_\mathrm{Kaon}>0.2$ are
chosen by optimizing the quantity $S/\sqrt{S+B}$, where $S$ and $B$
are the numbers of $B^+$ signal and background events, respectively.
After both requirements, a background reduction factor of
approximately four orders of magnitude is accomplished.

\begin{figure}
\centering
a)
\includegraphics[width=.43\textwidth]{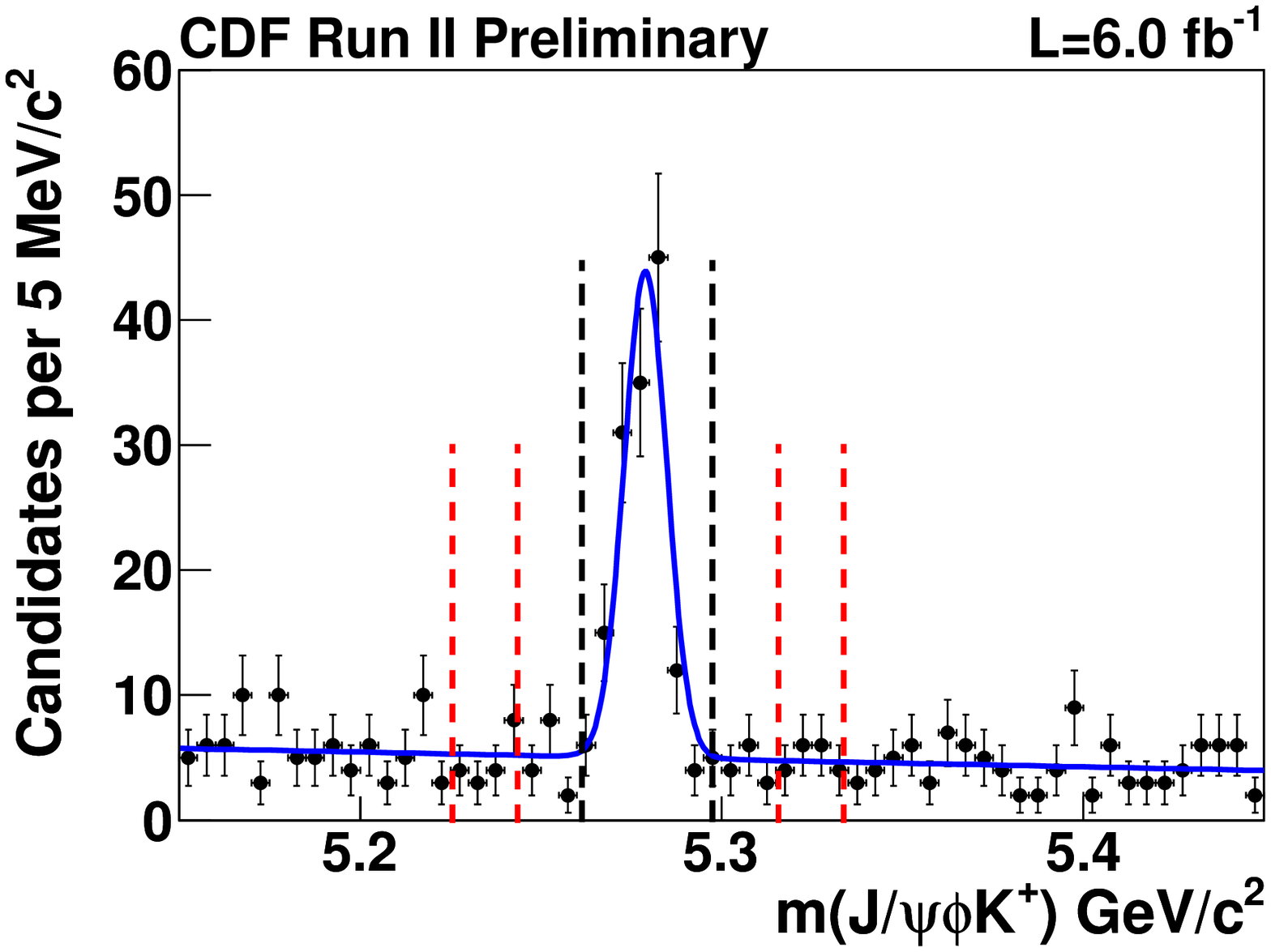}
\hspace{0.05\textwidth}
b)
\includegraphics[width=.43\textwidth]{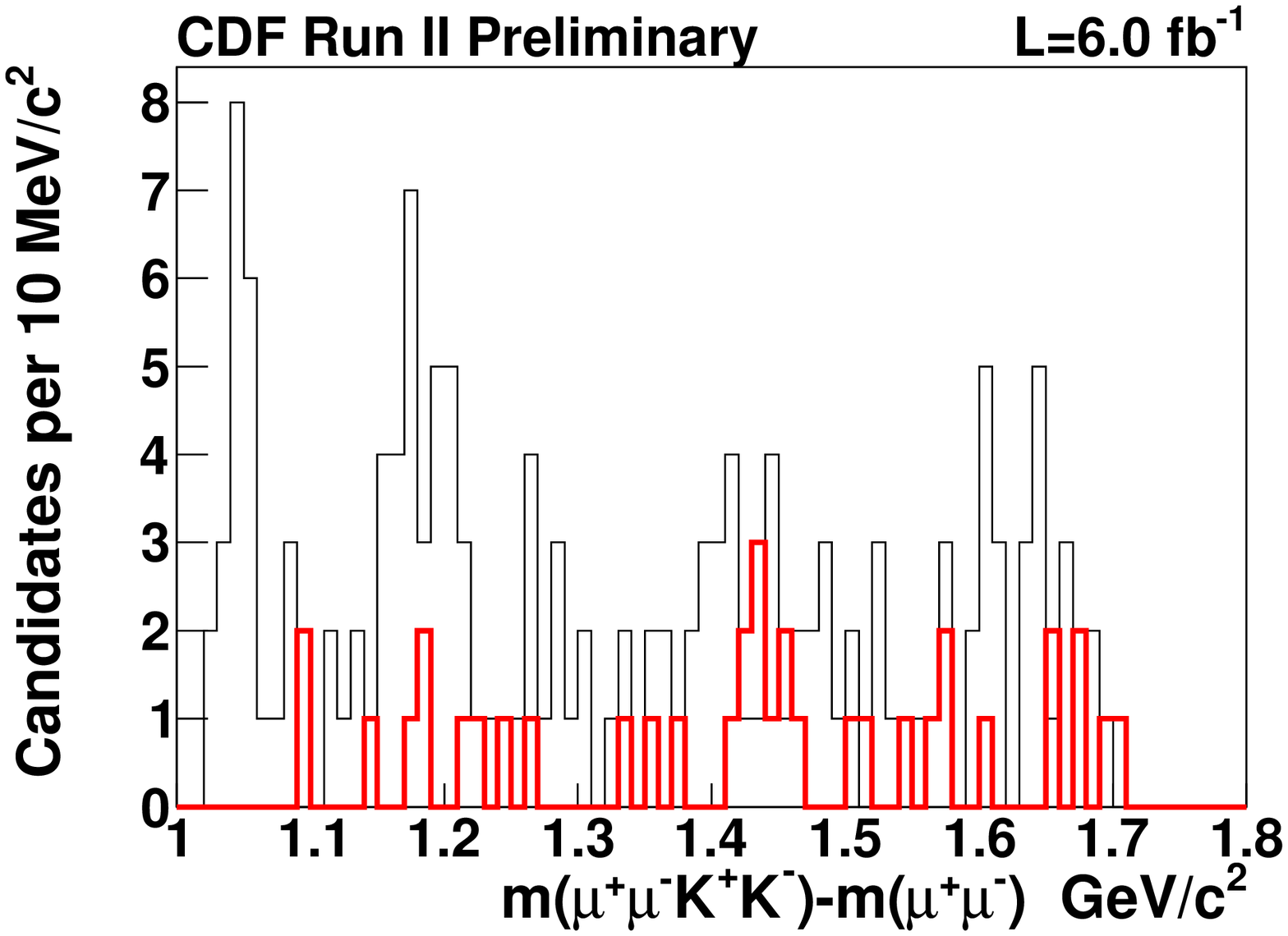}
\caption{(a) $J/\psi \phi K^+$ mass distribution with a fit to the
  data represented by the solid blue line. The vertical dashed black
  and red lines indicate the $B^+$ mass window and sidebands described
  in the text. (b) Mass difference distributions of the resulting
  $J/\psi \phi$ candidates from the $B^+$ mass window (black
  histogram) and sidebands (red histogram).}
\label{fig:B}
\end{figure}

\begin{figure}
\centering
a)
\includegraphics[width=.43\textwidth]{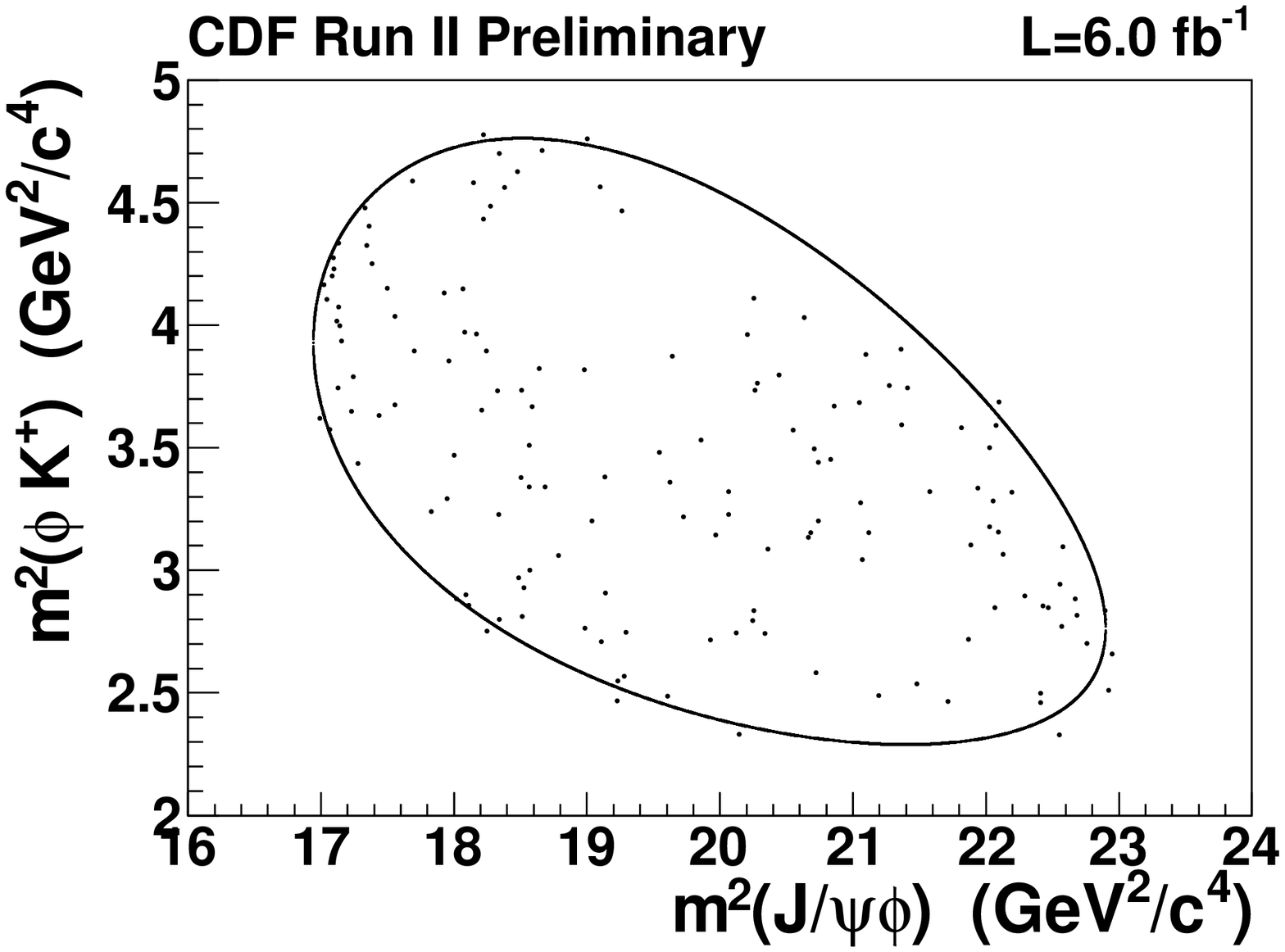}
\hspace{0.05\textwidth}
b)
\includegraphics[width=.43\textwidth]{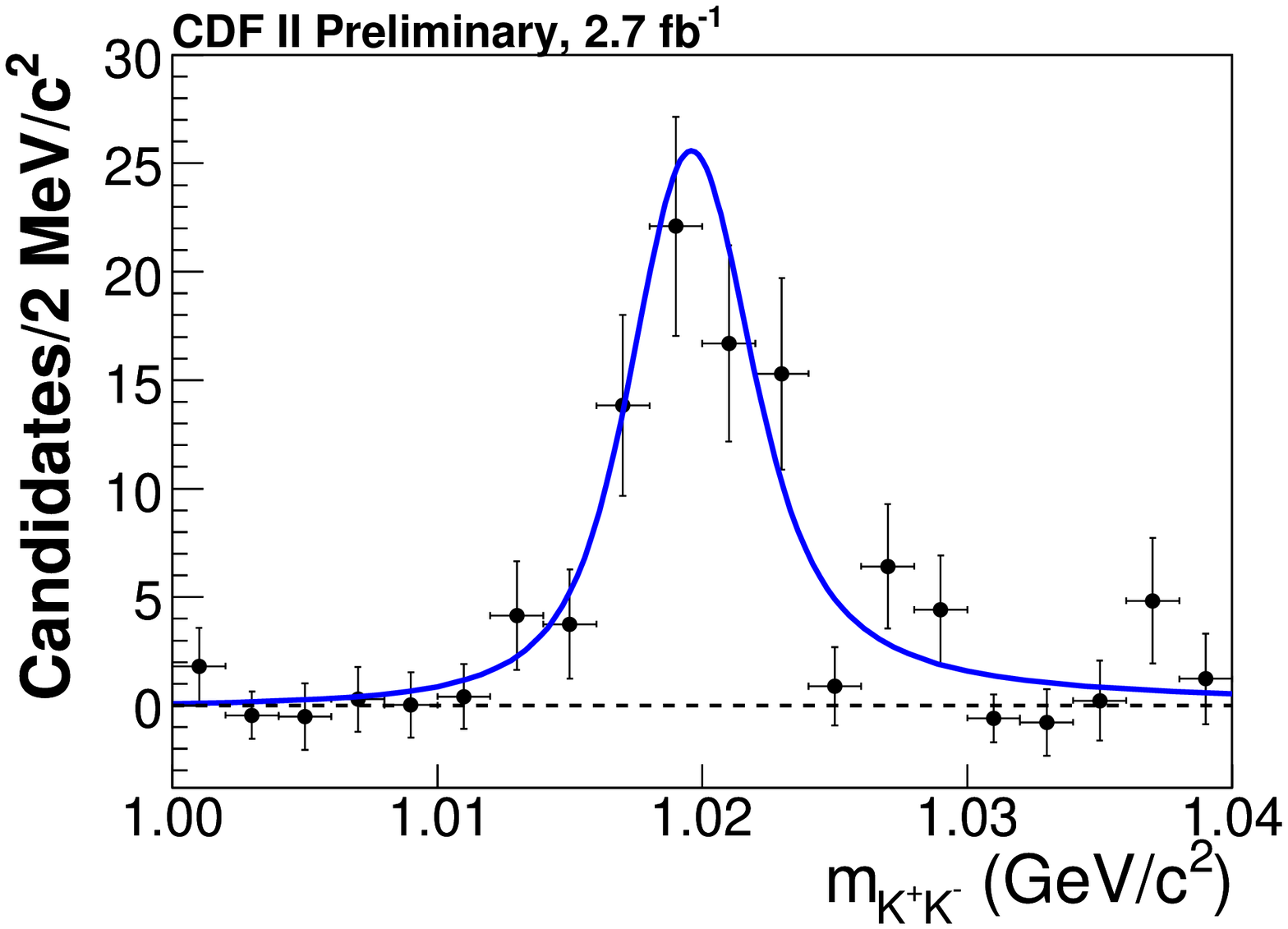}
\caption{(a) Dalitz plot of the final state $J/\psi \phi K^+$ in the
  $B^+$ mass window. The boundary shows the kinematically allowed
  region. (b) $K^+ K^-$ mass distribution with the fitted function
  (solid blue line) as described in the text.}
\label{fig:Valid}
\end{figure}

\begin{figure}
\centering
\includegraphics[width=.43\textwidth]{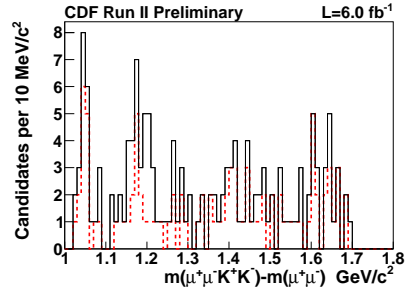}
\caption{Comparison between the $J/\psi \phi$ mass difference
  histograms of the dataset used for the updated analysis described in
  this write-up ($6.0\,\mathrm{fb}^{-1}$) in black and the one
  employed for the published $Y(4140)$ measurement
  ($2.7\,\mathrm{fb}^{-1}$) in dashed red.}
\label{fig:DatasetComp}
\end{figure}

Figure \ref{fig:B}(a) shows the resulting $J/\psi \phi K^+$ mass
spectrum. A fit with a Gaussian signal and a linear background
function yields $115 \pm 12$ signal events, corresponding to a 53\%
increase over the previous analysis. For the examination of the
$J/\psi \phi$ spectrum, only candidates within $\pm 3 \sigma$ ($\pm
17.7\,\mathrm{MeV/c^2}$) around the nominal $B^+$ mass are selected.
Furthermore, sideband events within $[-9 , -6]\sigma$ and $[+6 ,
+9]\sigma$ around the nominal $B^+$ mass are used to model the
combinatorial background in the $J/\psi \phi$ spectrum. Figure
\ref{fig:B}(b) shows the mass difference $\Delta M = m(\mu^+ \mu^- K^+
K^-) - m(\mu^+ \mu^-)$ distributions of the resulting $J/\psi \phi$
candidates. Whereas the $Y(4140)$ can be seen as narrow near-threshold
excess in the $B$ mass window, no such evidence is found from the $B$
mass sidebands. Figure \ref{fig:Valid}(a) shows the Dalitz plot of the
candidates from the $B$ mass window and figure \ref{fig:Valid}(b) the
$B$ sideband-subtracted $K^+ K^-$ mass spectrum without $\phi$ mass
window requirement. The fit function is a $P$-wave relativistic
Breit-Wigner convolved with a Gaussian to account for the detector
resolution. As there is no significant background contribution, the
$B^+ \to J/\psi K^+ K^- K^+$ final state is well described as $J/\psi
\phi K^+$. A comparison between the $J/\psi \phi$ mass difference
distributions of the dataset used for the updated analysis described
in this write-up, corresponding to an integrated luminosity of
$6.0\,\mathrm{fb}^{-1}$, and the one employed for the published
$Y(4140)$ measurement\cite{Evidence}, corresponding to
$2.7\,\mathrm{fb}^{-1}$, can be found in figure \ref{fig:DatasetComp}.

\section{Fits to Data and Significance Determination}

\begin{figure}
\centering
a)
\includegraphics[width=.43\textwidth]{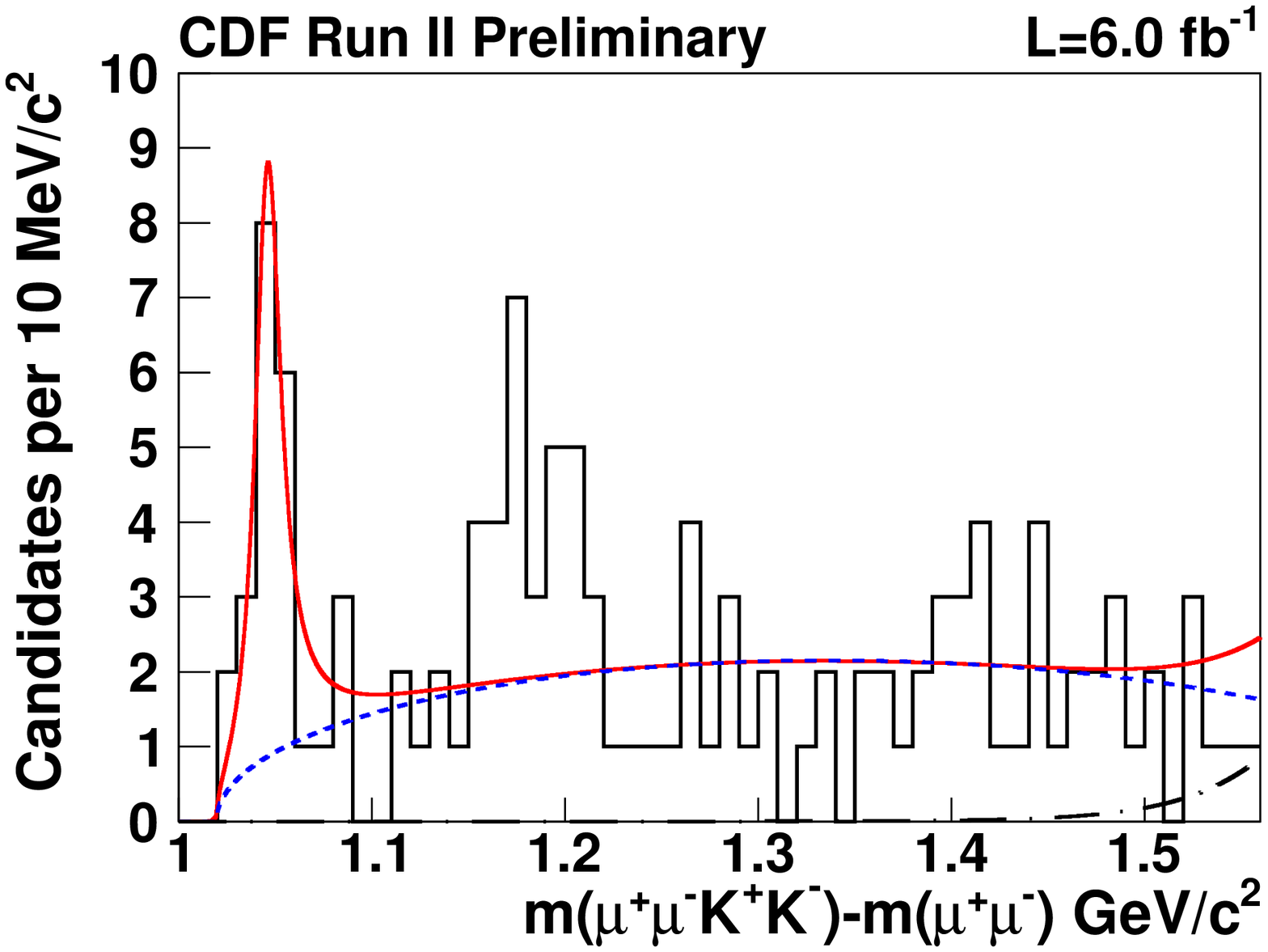}
\hspace{0.05\textwidth}
b)
\includegraphics[width=.43\textwidth]{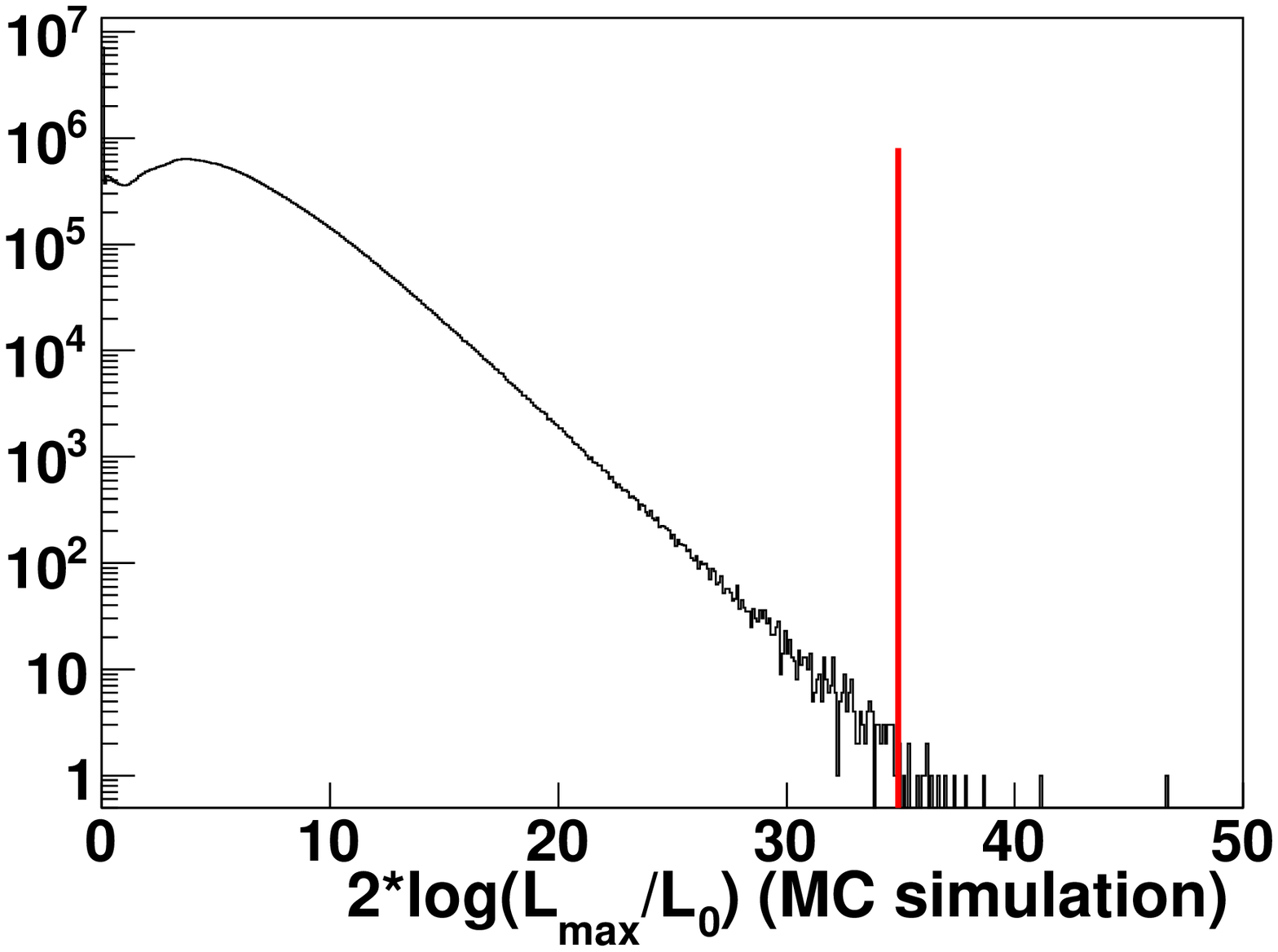}
\caption{(a) $J/\psi \phi$ mass difference distribution with a fit to
  the data represented by the solid red line. (b) $-2
  \ln{(\mathcal{L}_0/\mathcal{L}_{max})}$ distribution for 84 million
  simulation trials. The vertical red line indicates the value
  obtained in data.}
\label{fig:Y1}
\end{figure}

Figure \ref{fig:Y1}(a) shows the $\Delta M$ distribution from the $B$
mass window excluding events with $\Delta M > 1.56\,\mathrm{GeV/c^2}$
in order to avoid combinatorial backgrounds from misidentified $B_s^0
\to \psi(2S)\,\phi \to (J/\psi\,\pi^+\,\pi^-)\,\phi$ decays. An
unbinned maximum likelihood fit is performed, where the enhancement is
described by the convolution of an $S$-wave relativistic Breit-Wigner
function with a Gaussian resolution of $1.7\,\mathrm{MeV/c^2}$
obtained from Monte Carlo simulations, and the background is modeled
by three-body phase space. Including systematic uncertainties, which
are estimated by varying the fit model, the fit yields $19 \pm
6(\mathrm{stat}) \pm 3(\mathrm{syst})$ signal events. It returns a
mass of $m = 4143.4^{+2.9}_{-3.0}(\mathrm{stat}) \pm
0.6(\mathrm{syst})\,\mathrm{MeV}/c^2$ after including the world
average $J/\psi$ mass and a decay width of $\Gamma =
15.3^{+10.4}_{-6.1}(\mathrm{stat}) \pm
2.5(\mathrm{syst})\,\mathrm{MeV}/c^2$, both consistent with the values
from the published measurement\cite{Evidence}.  The observed width,
which is much larger than the resolution, suggests a strong decay for
the $Y(4140)$.  Furthermore, the relative branching fraction to the
nonresonant $B^+ \to J/\psi\,\phi\,K^+$ decay is measured as
$\frac{\mathcal{B}(B^+ \to Y(4140)\,K^+,\,Y(4140) \to
  J/\psi\,\phi)}{\mathcal{B}(B^+ \to J/\psi\,\phi\,K^+)} = 0.149 \pm
0.039(\mathrm{stat}) \pm 0.024(\mathrm{syst})$, where the relative
efficiency is determined to be $1.1$, using an $S$-wave relativistic
Breit-Wigner function with mean and width values determined from data
to represent the $Y(4140)$ structure and three-body phase space
kinematics for the nonresonant $B^+ \to J/\psi\,\phi\,K^+$ decay.

In order to estimate the probability of a creation of such a signal
due to background fluctuations, a large number of three-body phase
space $B^+$ decays are performed and the number of trials which
produce a signal with a log-likelihood ratio $-2
\ln{(\mathcal{L}_0/\mathcal{L}_{max})}$ of the null hypothesis fit and
the signal hypothesis fit larger than the value measured in data are
counted (see figure \ref{fig:Y1}(b)).  Thereby, the mass can be
anywhere in the considered $\Delta M$ window and the width has to be
larger than the detector resolution and smaller than
$120\,\mathrm{MeV}/c^2$. This procedure leads to a $p$-value of $2.3
\cdot 10^{-7}$, corresponding to a significance of the enhancement of
$5.0 \sigma$.

\begin{figure}
\centering
a)
\includegraphics[width=.43\textwidth]{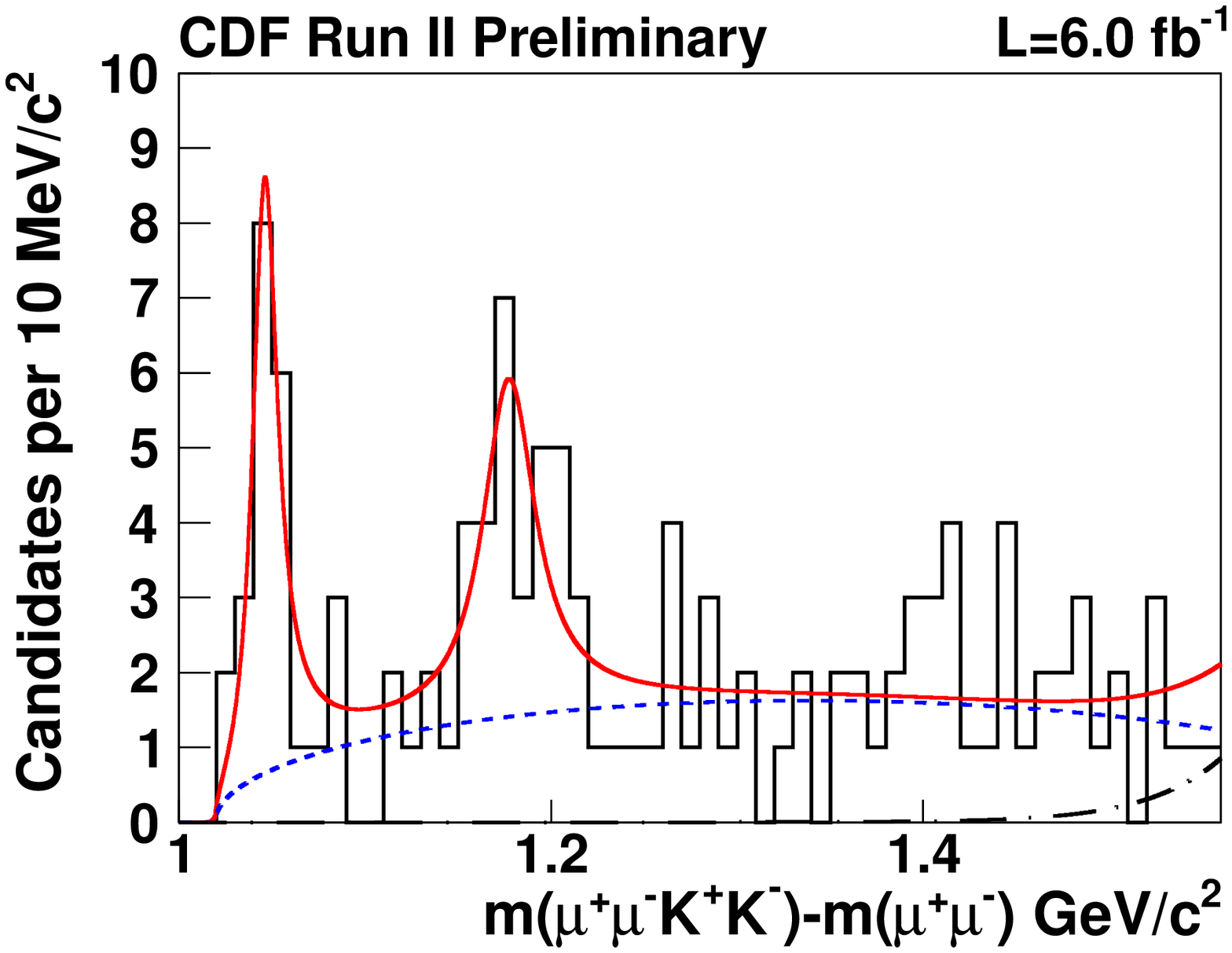}
\hspace{0.05\textwidth}
b)
\includegraphics[width=.43\textwidth]{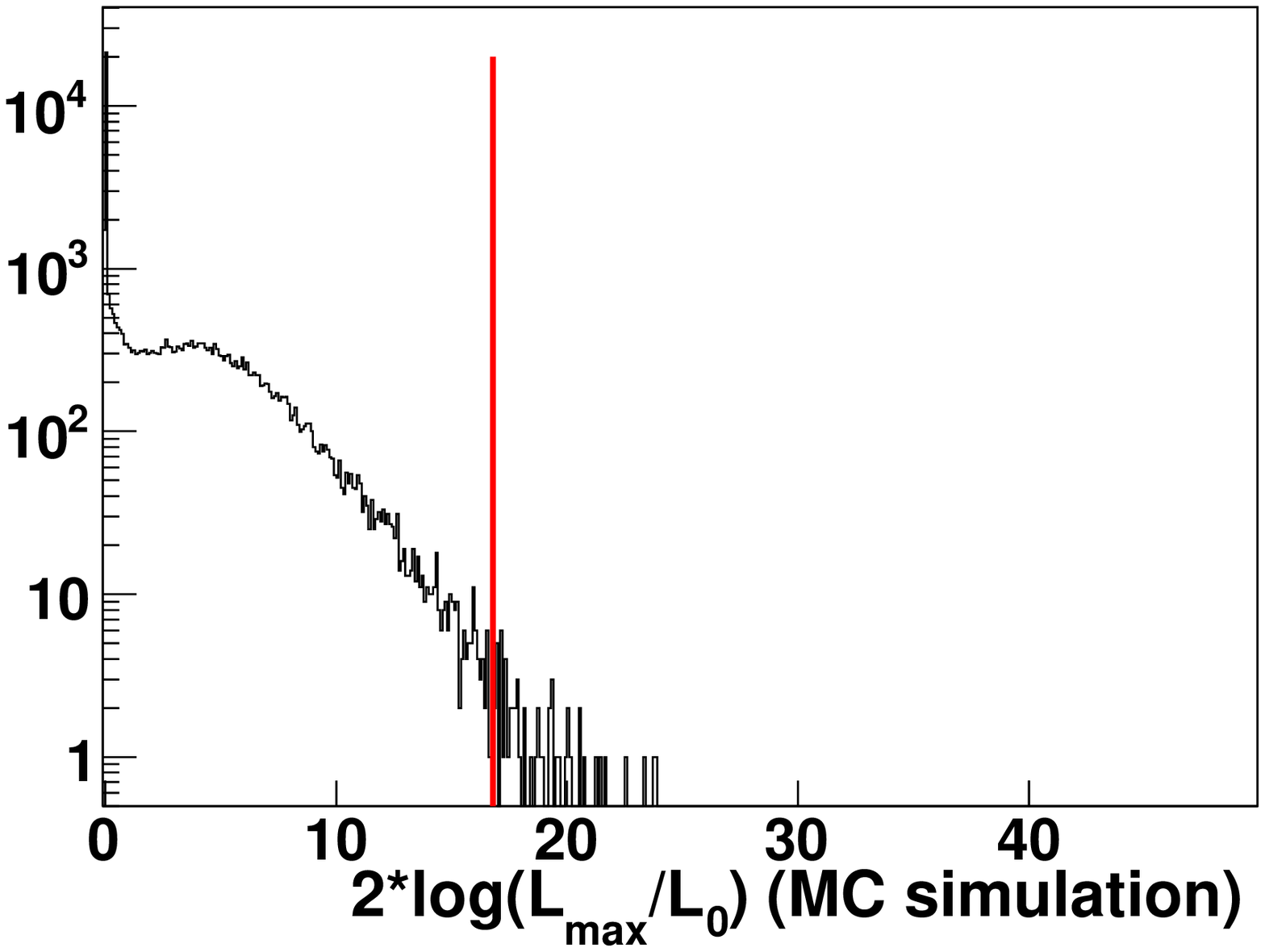}
\caption{(a) $J/\psi \phi$ mass difference distribution with a fit to
  the data, where an additional signal structure is included which is
  located about one pion mass higher than the $Y(4140)$. (b)
  Distribution of the log-likelihood ratio of the fit assuming only
  the $Y(4140)$ and the fit assuming two signal structures for the
  simulation trials. The vertical red line indicates the value
  obtained in data.}
\label{fig:Y2}
\end{figure}

In figure \ref{fig:Y1}(a), an additional excess above the background
appears at a mass of approximately $1.18\,\mathrm{GeV}/c^2$. With the
parameters of the $Y(4140)$ fixed to the values obtained from the fit
described above, another unbinned maximum likelihood fit assuming two
structures and the same background model as before is performed.
Thereby, the additional enhancement is described by the convolution of
an $S$-wave relativistic Breit-Wigner function with a Gaussian
resolution of $3.0\,\mathrm{MeV/c^2}$ obtained from Monte Carlo
simulations.  The measured data distribution together with the fit
projection can be found in \ref{fig:Y2}(a). The fit returns a yield of
$22 \pm 8$ signal events, a mass, after including the world average
$J/\psi$ mass, of $m = 4274.4^{+8.4}_{-6.7}\,\mathrm{MeV}/c^2$ and a
decay width of $\Gamma = 32.3^{+21.9}_{-15.3}\,\mathrm{MeV}/c^2$. Just
like in the $Y(4140)$ case, the statistical significance of the
additional excess is determined by simulations, where the
log-likelihood ratio of the fit assuming only the $Y(4140)$ and the
fit assuming two signal structures is calculated (see figure
\ref{fig:Y2}(b)). This leads to a $p$-value of $1.1 \cdot 10^{-3}$,
corresponding to a significance of $3.1 \sigma$.

After the first confirmation of Belle's $X(3872)$, including the
determination of its allowed quantum numbers and the most precise mass
measurement,\cite{X} CDF keeps contributing to the field of exotic
$X,Y,Z$ states with this recent observation of a narrow structure,
called $Y(4140)$, near the $J/\psi\,\phi$ threshold in exclusive $B^+
\to J/\psi\,\phi\,K^+$ decays.

\end{document}